\documentclass[twocolumn,showpacs,preprintnumbers,amsmath,amssymb]{revtex4}
\usepackage{dcolumn}
\usepackage{bm}
\usepackage{graphicx}

\usepackage{hyperref}
\hypersetup{colorlinks,
linkcolor=blue,
filecolor=green,
urlcolor=blue,
citecolor=blue}

\def\bk{{\bf k}}
\def\wk{\omega_k}

\def\akj{a_{\bk j}}
\def\akjd{a_{\bk j}^\dagger}

\def\bD{{\bf D}}
\def\bd{{\bf d}}
\def\br{{\bf r}}
\def\bp{{\bf p}}
\def\k0{k_0}

\def\ekj{\hat{e}_{\bk j}}

\begin{document}

\title{Enhanced resonant force between two entangled identical atoms in a photonic crystal}

\author{Roberta Incardone${}^{1}$}
\altaffiliation[Present address: ]{Max-Planck-Instit\"{u}t f\"{u}r Intelligente Systeme, Heisenbergst. 3, D-70569 Stuttgart, Germany and IV Institut f\"{u}r Theoretische Physik, Universit\"{a}t Stuttgart, Pfaffenwaldring 57,  D-70569 Stuttgart, Germany}
\author{Taku Fukuta${}^{2}$}
\author{Satoshi Tanaka${}^{2}$}
\author{Tomio Petrosky${}^{3}$}
\author{Lucia Rizzuto${}^{1}$}
\author{Roberto Passante${}^{1}$}
\affiliation{
${}^{1}$ Dipartimento di Fisica e Chimica, Universit\`{a} degli Studi di Palermo and CNISM, Via Archirafi 36, I-90123 Palermo, Italy \\
${}^{2}$ Department of Physical Science, Osaka Prefecture
University, Gakuen-cho 1-1, Sakai 599-8531, Japan \\
${}^{3}$ Center for Studies in Statistical Mechanics and Complex
Systems, The University of Texas at Austin, Austin, TX 78712 USA}


\begin{abstract}
We consider the resonant interaction energy and force between two identical atoms, one in an excited state and the other in the ground state, placed inside a photonic crystal. The atoms, having the same orientation of their dipole moment, are supposed prepared in their symmetrical state and interact with the quantum electromagnetic field. We consider two specific models of photonic crystals: a one-dimensional model and an isotropic model. We show that in both cases the resonant interatomic force can be strongly enhanced by the presence of the photonic crystal, as a consequence of the modified dispersion relation and density of states, in particular if the transition frequency of the atoms is close to the edge of a photonic gap. Differences between the two models considered of photonic crystal are discussed in detail, as well as comparison with the analogous system of two impurity atoms in a quantum semiconductor wire. A numerical estimate of the effect in a realistic situation is also discussed.

\end{abstract}

\maketitle

\section{\label{sec:1}Introduction}

Dispersion and resonant interactions are intermolecular interactions between neutral atoms or molecules due to their common interaction with the quantum electromagnetic radiation field \cite{Buhmann}. Dispersion or van der Waals interactions play a role in many different physical phenomena \cite{DMRR11}, as well as in chemistry and biology \cite{Frolich72}, and direct and indirect measurements of such interactions have been recently done, even in the so-called Casimir-Polder regime \cite{PJCKMS12,BVCLB13}. Resonant interactions occur when one or more molecules are in their excited state and the interaction, being involved the exchange of real photons between the atoms, can be of very long range \cite{CP98,Salam10,PT93,BLMN03,PPR05}. The resonant interatomic force is also related to the resonant energy transfer between molecules \cite{Salam08}, which has been recently guessed to play a role in coherent biological phenomena such as photosynthesis \cite{Scholes03,MRLAG08,FOC10} or interactions between macromolecules \cite{PP13}.

In this paper we consider the resonant interaction between two identical atoms/molecules, one excited and the other in the ground state, prepared in an entangled symmetrical state, when the two atoms are placed inside a structured environment such as a photonic crystal. We first consider an isotropic model of the photonic crystal and then a one-dimensional (1D) model. The photonic crystal strongly modifies the dispersion relation and density of states for the photons (see for example \cite{LNNB00,AKP04}, and references therein). Strong modifications of the spontaneous emission rate \cite{JQ94,JW91,NFA07}, Lamb shift \cite{WKG04} and resonant dipole-dipole energy transfer \cite{BLM97,HCJLW12} for atoms in a photonic crystal have been considered in the literature. Motivation of this work is to investigate the possibility of modification, enhancement in particular, of the resonant interatomic force due to the modified photonic dispersion relation and density of states in the presence of the photonic crystal. A similar effect has been recently shown for the electronic Casimir-Polder interaction between two impurities in a one-dimensional semiconductor wire \cite{TPFP13}. Also, strong increase of van der Waals and Casimir-Polder interactions for neutral atoms in the vicinity of transmission lines has been recently discussed in the literature \cite{SMK13}. In the present paper we consider the case when the transition frequency of the two atoms is very close to the edge of the band gap of the photonic crystal, where the density of states is very large (van Hove singularity). For both models of photonic crystal considered, we find a strong increase of the interatomic resonant force compared to the case of atoms in the free space, although the distance dependence of the force is the same as in the free space. We also show that analogous results are expected in different physical one-dimensional system, for example the case of two impurity atoms in a quantum semiconductor wire. Limits of our approach are also discussed.

This paper is organized as follows. In Section \ref{sec:2} we introduce our model of the two identical atoms, one excited and the other in the ground state and prepared in a symmetrical state, with the field in its vacuum state. We derive a closed equation for the resolvent of the system, and the energy shift of the atoms due to their interaction with the field is obtained from the poles of the resolvent. Up to this point, our results are general and valid for any field dispersion relation. In Section \ref{sec:3} we introduce our specific models (1D and isotropic 3D models) for the photonic crystal and the relative dispersion relation; after solving by iteration the equation for the resolvent, we obtain the interatomic potential energy in both cases, and then the interatomic force, when the atomic transition frequency is close to the edge of the photonic band gap. Section \ref{sec:4} is devoted to our conclusive remarks.

\section{\label{sec:2}The model}

We consider two identical atoms A and B separated by a distance $r$ and interacting with the quantum radiation field in the multipolar coupling scheme and within dipole approximation. The two atoms, that for simplicity we consider as two-level systems, are inside an isotropic photonic crystal, as discussed later on.

The Hamiltonian of our system, in the multipolar coupling scheme (three-dimensional case)  \cite{CPP95}, is

\begin{eqnarray}
H &=& H_0 + H_I \, ,\nonumber \\
H_0 &=& H_A + H_B + \sum_{\bk j} \hbar \wk \akjd \akj  \, , \nonumber \\
H_I &=& - {\boldsymbol\mu}_A \cdot \bD_\perp (0)  - {\boldsymbol\mu}_B \cdot \bD_\perp (\br ) \, ,
\label{eq:1}
\end{eqnarray}
where A, B indicate the two atoms with position $0$ and $\br$ respectively ($H_A$ and $H_B$ are their Hamiltonian), ${\boldsymbol\mu}_A = e\br_A$ and
${\boldsymbol\mu}_B=e\br_b$ are their dipole moment operators. $\bD_\perp(\br )$ is the transverse displacement field operator, given by
\begin{equation}
\bD_\perp(\br ) = i\sum_{\bk j} \ekj  \left( \frac{2\pi \hbar \wk}V \right)^{1/2} \left( \akj e^{i\bk \cdot \br} - \akjd e^{-i\bk \cdot \br} \right) \, ,
\label{eq:1a}
\end{equation}
where $\akj, \, \akjd$ are bosonic annihilation and creation operators, $\ekj$ are polarization unit vectors and $V$ is the quantization volume.

In the 1D case, the interaction term of the Hamiltonian is conveniently written as
$H_I = -  \bp_A \cdot \bd_\perp (0)  - \bp_B \cdot \bd_\perp (r )$ where $\bp_{A,B}$ is the atom-field coupling constant, proportional to the atomic dipole moment operator, and the 1D transverse displacement field is given by
\begin{equation}
\bd_\perp(r ) = i\sum_{k j} \hat{e}_{kj}  \left( \frac{2\pi \hbar \wk}\ell \right)^{1/2} \left( a_{kl} e^{i k r} - a_{kl}^\dagger e^{-i k r} \right) \, ,
\label{eq:1b}
\end{equation}
$\ell$ being the quantization length.

The relation between the photon frequency $\wk$ and wavenumber $k$ depends on the specific model for the photonic crystal, as we shall discuss later on.

We assume the two atoms identical and the system prepared in the following symmetrical state

\begin{equation}
\mid i \rangle = \frac 1{\sqrt{2}} \left( \mid g_A, e_B; 0_{\bk j} \rangle + \mid e_A, g_B; 0_{\bk j} \rangle \right) \, ,
\label{eq:2}
\end{equation}
where $g$ ($e$) indicates the ground (excited) state of the atom, and $0_{\bk j}$ the photon vacuum. The energy separation of the two atomic levels is $E_i$. In such a state the atomic excitation is delocalized among the two atoms. This state is also called a superradiant state, because in the Dicke model its decay rate is larger than that of the individual atoms, yielding a collective spontaneous decay. Another possible state where the atomic energy is delocalized is the antisymmetric combination, which in the same model is a subradiant state \cite{Dicke54,EGJ13}.
In this paper we consider only the case in which the two atoms are prepared in the symmetrical state \eqref{eq:2}.
Recently, different methods for entangling systems of two-level atoms in a structured environment such as a photonic crystal have been proposed \cite{KB00}. In particular it has been discussed how photonic bandgap materials can be used to preserve the maximally symmetric entangled state \cite{BLMC08, WJZHZ11} .
Also, it has been shown that symmetrical state can be
easily realized considering two superconducting qubits in 1-D trasmission line, where the photon-mediated interaction between
the two artificial atoms leads to correlated states
\cite{LSLFWB13,LFLSBW13}.

The interaction energy between the two atoms can be obtained using the resolvent formalism: the discrete level energies are given by the poles of the resolvent \cite{CTDRG92}. In a quasi static approach, the force can be then obtained as the negative of the derivative of the energy with respect to the distance between the atoms. We assume our state $\mid i \rangle$ as a stable state, and thus our results are valid only for times shorter than its spontaneous decay time.

First step is to obtain a closed set of equations for the matrix elements of the resolvent $G(z)=(z-H)^{-1}$. The energy shift of the system due to the interaction can be obtained from the poles of the resolvent. We take the expectation value of the relation $(z-H_0)G(z)=1+H_I G(z)$ on the state $\mid i \rangle$ given by Eq. \eqref{eq:2}, obtaining
\begin{equation}
(z-E_i)G_{ii}(z)=1+\sum_\ell \langle i \mid H_I  \mid \ell \rangle G_{\ell i}(z) \, ,
\label{eq:3}
\end{equation}
where $G_{\ell i}(z)=\langle \ell \mid G(z) \mid i \rangle$ and $\ell$ denotes a complete set of intermediate states. From \eqref{eq:1}, it is immediate to see that possible intermediate states are $\mid g_A, g_B; 1_{\bk j} \rangle$ and $\mid e_A, e_B; 1_{\bk j} \rangle$.
The latter is a virtual state, and it is possible to show that its contribution to the interatomic energy can be neglected at large interatomic distances, such that $E_i r /\hbar c \gg 1$.
In fact, at the second order in perturbation theory and for atoms in the vacuum space, in \cite{CP98,Salam10} it is shown that the virtual intermediate state
$\mid e_A, e_B; 1_{\bk j} \rangle$ gives a contribution proportional to the following integral over $k$
\begin{equation}
\frac 1r \int_0^\infty dk \frac{\sin kr}{k_0+k}= \frac 1r f(k_0r)  \, ,
\label{eq:3a}
\end{equation}
while the corresponding term from the real intermediate state $\mid g_A, g_B; 1_{\bk j} \rangle$ is
\begin{equation}
\frac 1r \int_0^\infty dk \frac{\sin kr}{k_0-k}= \frac 1r f(k_0r) - \frac {\pi \cos k_0r}r \, ,
\label{eq:3b}
\end{equation}
where
$f(z)= \text{Ci}(z)\sin (z)- \text{si}(z) \cos (z)$ is the auxiliary function of the sine and cosine integral functions \cite{AS65}. Because $f(k_0r) \sim 1/(k_0r)$ for $k_0r  \gg 1$, the contribution \eqref{eq:3a} becomes negligible compared to the contribution \eqref{eq:3b} in the long-distance regime $k_0r \gg 1$ we are considering.

From now on we consider this large distance limit only and thus we neglect the virtual intermediate states in \eqref{eq:3}.
Using \eqref{eq:1} and \eqref{eq:2}, Eq.  \eqref{eq:3} yields (3D case)
\begin{widetext}
\begin{equation}
(z-E_i) G_{ii}(z)= 1-i\sum_{\bk j}\sqrt{\frac{\pi\hbar\wk}V}\left( \ekj \cdot {\boldsymbol\mu}_A^{eg} + \ekj \cdot {\boldsymbol\mu}_B^{eg} e^{i\bk \cdot \br} \right) \langle g_A, g_B; 1_{\bk j} \mid G(z) \mid i \rangle \, ,
\label{eq:4}
\end{equation}
\end{widetext}
where ${\boldsymbol\mu}_{A(B)}^{eg}$ is the matrix element of the dipole moment of atom $A(B)$.
In obtaining Eq. \eqref{eq:4} we have assumed a three-dimensional case, and thus it is valid for the isotropic photonic crystal we shall discuss in the next Section. An analogous expression is obtained for the 1D case, that we do not report here for brevity.
We now assume that the two identical atoms $A$ and $B$ also have the same matrix element of the dipole moment ${\boldsymbol\mu}^{eg}= {\boldsymbol\mu}^{eg}_A={\boldsymbol\mu}^{eg}_B$, i.e. that they are directed along the same direction. This hypothesis is essential to obtain a closed equation for the resolvent, as it will become evident later on in this Section. Thus Eq. \eqref{eq:4} becomes
\begin{eqnarray}
(z-E_i) G_{ii}(z)&=& 1-i\sum_{\bk j} \sqrt{\frac{\pi\hbar\wk}V} \ekj \cdot {\boldsymbol\mu}^{eg} \nonumber \\
&\times& \! \left( 1+ e^{i\bk \cdot \br} \right) \langle g_A, g_B; 1_{\bk j} \mid G(z) \mid i \rangle \, .
\label{eq:5}
\end{eqnarray}

In order to obtain a closed equation for $G_{ii}(z)$, we can write the matrix element of the relation $(z-H_0)G(z)=1+H_I G(z)$ for the resolvent, between the states $\mid g_A, g_B; 1_{\bk j} \rangle$ and $\mid i \rangle$ (with the hypothesis of identical matrix elements of the dipole moments of the two atoms, as before),
\begin{widetext}
\begin{equation}
(z-\hbar \wk ) \langle g_A, g_B; 1_{\bk j} \mid G(z) \mid i \rangle = i\sqrt{\frac{2\pi\hbar\wk}V} \ekj \cdot {\boldsymbol\mu}^{ge} \left[
\langle e_A, g_B; 0_{\bk j} \mid G(z) \mid i \rangle + e^{-i\bk \cdot \br} \langle g_A, e_B; 0_{\bk j} \mid G(z) \mid i \rangle \right] \, .
\label{eq:6}
\end{equation}
\end{widetext}

In this Section we do not specify a particular dispersion relation between $\wk$ and $\bk$. First we only assume that, for the three-dimensional case, it is isotropic, that is that $\wk$ depends only on the modulus $k$ of the wavevector, and not from its direction \cite{AKP04}. Then, we shall specify our calculation also to the case of a one-dimensional photonic crystal.
Substituting \eqref{eq:6} into \eqref{eq:5}, we obtain
\begin{widetext}
\begin{eqnarray}
(z-E_i) G_{ii}(z) &=& 1+ \frac {\sqrt{2}\pi \hbar}V\sum_{\bk j}  \frac \wk{z-\hbar \wk } \left( \ekj \cdot {\boldsymbol\mu}^{eg} \right) \left( \ekj \cdot {\boldsymbol\mu}^{ge} \right)
\nonumber \\
&\times& \! \left(  e^{-i\bk \cdot \br} \langle g_A, e_B; 0_{\bk j} \mid G(z) \mid i \rangle + e^{i\bk \cdot \br} \langle e_A, g_B; 0_{\bk j} \mid G(z) \mid i \rangle + \sqrt{2} G_{ii}(z) \right) \, .
\label{eq:7}
\end{eqnarray}
\end{widetext}

Going to the continuum limit  $\sum_\bk \rightarrow V/(2\pi )^3 \int \! dk k^2 d\Omega$, after performing the polarization sum using $\sum_j (\ekj)_m (\ekj)_n = \delta_{mn}-\hat{k}_m \hat{k}_n$, we can perform the angular integration
\begin{eqnarray}
&\ & \int \! d\Omega \left( \delta_{mn} -\hat{k}_m \hat{k}_n \right) e^{\pm i\bk \cdot \br} \nonumber \\
&=& \frac 1 {k^2} \left( -\nabla^2 \delta_{mn}+\nabla_m \nabla_n \right)
\int \! d\Omega e^{\pm i\bk \cdot \br}
\nonumber \\
&=&  \frac {4\pi}{k^3}\left( -\nabla^2 \delta_{mn}+\nabla_m \nabla_n \right) \frac {\sin kr}r \, ,
\label{eq:8}
\end{eqnarray}
where the differential operators in \eqref{eq:8} are with respect to $\br$. This allows us to combine the first two terms in the second line of Eq. \eqref{eq:7} in terms of the matrix element $G_{ii}(z)$ only.
Eq. \eqref{eq:7} thus yields (repeated indices are implicitly summed over)
\begin{widetext}
\begin{eqnarray}
(z-E_i) G_{ii}(z) &=& 1+ G_{ii}(z) \frac \hbar{4\pi^2}\int_0^\infty \! dk k^2 \frac \wk {z-\hbar \wk } \int \! d\Omega \left( \mu^{eg} \right)_m  \left( \mu^{ge} \right)_n \left( \delta_{mn}-\hat{k}_m \hat{k}_n \right)
\nonumber \\
&+& G_{ii}(z) \frac \hbar\pi \left( \mu^{eg} \right)_m  \left( \mu^{ge} \right)_n \left( -\nabla^2 \delta_{mn}+\nabla_m \nabla_n \right)
\int_0^\infty \! dk \frac \wk {z-\hbar \wk } \frac {\sin kr}{kr}  \, .
\label{eq:9}
\end{eqnarray}
\end{widetext}

Eq. \eqref{eq:9} is valid for a 3D case, independently from a specific dispersion relation.
It is an equation for the resolvent $G_{ii}(z) $, from which one can find its poles, giving the discrete energy levels of the coupled atoms-field system. The second term on the right-hand side of the first line of \eqref{eq:9} does not depend on $r$. Therefore this term gives an energy shift $\Delta$ that does not depend on the distance $r$ between the two atoms, and thus it does not contribute to their interaction energy. For this reason, in the next Section we will not consider it explicitly. The distance-dependent part of the energy is thus given by the solutions $z$ of the implicit equation

\begin{eqnarray}
z &=& E_i +\Delta + \frac \hbar\pi \left( \mu^{eg} \right)_m  \left( \mu^{ge} \right)_n \nonumber \\
&\times& \! \! \left( -\nabla^2 \delta_{mn}+\nabla_m \nabla_n \right)
\int_0^\infty \! dk \frac \wk {z-\hbar \wk } \frac {\sin kr}{kr}  \, ,
\label{eq:10}
\end{eqnarray}
where
\begin{eqnarray}
\Delta &=& \frac \hbar{4\pi^2}\int_0^\infty \! dk k^2 \frac \wk {z-\hbar \wk } \nonumber \\
&\times&\int \! d\Omega \left( \mu^{eg} \right)_m  \left( \mu^{ge} \right)_n \left( \delta_{mn}-\hat{k}_m \hat{k}_n \right)
\label{eq.10n}
\end{eqnarray}
is a distance-independent energy shift (Lamb shift).

Equation \eqref{eq:10} is exact within our isotropic model, except for our approximation of neglecting the virtual intermediate states compared to the real ones (long-distance hypothesis), and no perturbative expansion has been so far used.

Following the same procedure yielding \eqref{eq:7}, it is possible to obtain the analogous equation for the one-dimensional case, using the interaction Hamiltonian we have given before Eq. \eqref{eq:1b}, with the expression \eqref{eq:1b} for the 1D transverse displacement field. In this case, the equation for the resolvent analogous to our previous Eq. \eqref{eq:7} is
\begin{widetext}
\begin{eqnarray}
(z-E_i) G_{ii}(z) &=& 1+ \frac {\sqrt{2}\pi \hbar}\ell \sum_{kj} \left( \hat{e}_{kj} \cdot \bp^{eg} \right) \left( \hat{e}_{kj} \cdot \bp^{ge} \right)
\nonumber \\
&\times& \! \frac \wk{z-\hbar \wk } \left( e^{-ikx} \langle g_A, e_B; 0_{kj} \mid G(z) \mid i \rangle + e^{ikx} \langle e_A, g_B; 0_{kj} \mid G(z) \mid i \rangle+ \sqrt{2} G_{ii}(z) \right) \, ,
\label{eq:10a}
\end{eqnarray}
\end{widetext}
where $x$ is the distance between the two atoms (along the direction considered in our 1D model). In the continuum limit, $\sum_k \rightarrow (\ell / 2\pi )\int_{-\infty}^{\infty}dk$, we obtain
\begin{widetext}
\begin{eqnarray}
(z-E_i) G_{ii}(z) &=& 1+ \frac{\hbar}{\sqrt{2}} \left( \mid \bp^{eg} \mid^2 - \mid \bp^{eg}_x \mid^2  \right)
\nonumber \\
&\times& \! \int_{-\infty}^\infty \! dk \frac \wk{z-\hbar \wk } \left( e^{-ikx} \langle g_A, e_B; 0_{kj} \mid G(z) \mid i \rangle + e^{ikx} \langle e_A, g_B; 0_{kj} \mid G(z) \mid i \rangle + \sqrt{2} G_{ii}(z) \right) \, .
\label{eq:10b}
\end{eqnarray}
\end{widetext}
The two integrals over $k$ containing the exponential factors are equal, and this allows to close the equation for $G_{ii}(z)$, and obtain its poles by the implicit equation
($\wk = \omega_{-k}$)
\begin{eqnarray}
z &=& E_i +\Delta_1 + 2\hbar \left( \mid \bp^{eg} \mid^2 - \mid \bp^{eg}_x \mid^2 \right) \nonumber \\
&\times& \! \!
\int_{0}^\infty \! dk \frac \wk {z-\hbar \wk }\cos(kx)  \, ,
\label{eq:10c}
\end{eqnarray}
where
\begin{equation}
\Delta_1 = 2\hbar \left( \mid \bp^{eg} \mid^2 - \mid \bp^{eg}_x \mid^2 \right) \int_{0}^\infty \! dk \frac \wk {z-\hbar \wk }
\label{eq:10d}
\end{equation}
is an energy shift independent from the interatomic distance, and thus not contributing to the force between the atoms.

Here we would like to compare the present result with that obtained in a one-dimensional electronic system composed of a quantum wire with two identical impurities \cite{Tanaka07}, where the pole is obtained by solving the following equation
\begin{equation}\label{El:disp}
z=E_0+{g^2 B^2\over 2\pi}\int_{-\pi}^\pi {1\pm \cos(k x) \over z-\hbar\omega_k} dk \;,
\end{equation}
where the sign plus or minus is for the symmetrical or antisymmetrical combination of the impurity states, respectively.

In Eq.(\ref{El:disp}), $E_0$ is the bare impurity energy, $B$ is a half bandwidth of the one-dimensional electronic band, and the energy dispersion is given by
\begin{equation}
\hbar\omega_k=-B\cos k \;.
\end{equation}
The similarity between Eqs.(\ref{eq:10c}) and (\ref{El:disp}) is obvious so that the present results are ubiquitously to be seen in different real physical systems.

\section{\label{sec:3}Resonant interatomic force in the photonic crystal}

We now specify specific models of photonic crystals and use the relative dispersion relations. The basic model we use, that has been largely used in the literature \cite{JQ94,JW91}, is a one-dimensional model with a periodic array of dielectric plane slabs of width $2a$ in vacuum, separated by a distance b. $L=2a+b$ is the periodicity of the crystal. The dielectric is assumed to have a real and frequency independent refractive index $n$. The dispersion relation for this 1D model is well known. In the isotropic model, it is assumed valid for any direction of the wave vector. This dispersion relation is particularly simple when $b=2na$ (we refer to \cite{JW91} for more details on this model of photonic crystal). The dispersion relation in this case is
\begin{equation}
\wk = \frac c{4na} \arccos \left[ \frac {4n\cos (kL) +(1-n)^2}{(1+n)^2} \right] \, .
\label{eq:11}
\end{equation}

The photonic crystal has photonic band gaps at wavenumber $k=q\pi/L$, with $q$ a positive integer number, as it can be obtained from Eq. \eqref{eq:11}.
We respectively denote with $\omega_v$ and $\omega_c$ the frequency of the lower and upper edge of the first photonic gap ($q=1$). Near the lower and upper edges of the gap, the dispersion relation can be expanded in powers of $k-k_0$ keeping quadratic terms only, obtaining the so-called effective mass approximation. Below the gap, this approximation gives
\begin{equation}
\wk = \omega_v - A (k-k_0)^2 \, ,
\label{eq:12}
\end{equation}
and above the gap,
\begin{equation}
\wk = \omega_c + A (k-k_0)^2 \, ,
\label{eq:13}
\end{equation}
where $k_0$ is the wave vector corresponding to the first gap and A is a positive constant. In this approximation and for frequencies above the gap and not too far from it, the density of photon states inside the crystal is
\begin{equation}
\rho (\wk ) = \frac {k_0^2}{\sqrt{A}} \frac {2\pi}{\sqrt{\wk - \omega_c}} \, .
\label{eq:14}
\end{equation}
A similar expression holds below the gap. For frequencies $\omega_v < \wk < \omega_c$, that is inside the gap, the density of states vanishes and photons in this frequency interval cannot propagate in the crystal.
Eq. \eqref{eq:14} shows that the density of states is very large near the edges of the photonic band gap, and it becomes singular at the edges (Van Hove singularity). This gives a strong coupling with field modes with frequencies in the proximity of the gap.

We will now consider two specific cases: a 1D case where we use relations \eqref{eq:12}, \eqref{eq:13}, \eqref{eq:14} for a one-dimensional crystal (we will discuss later on about validity and limits of the 1D approximation), and an isotropic 3D case where these relations are assumed valid for any direction of the wavevector $\bk$.

We first consider the isotropic crystal case. In order to obtain the interaction energy between the two atoms in the isotropic crystal, as a function of their distance, and then the force in the quasi-static approach, we need to solve Eq. \eqref{eq:10}, where the function $\wk /(z-\hbar \wk )$ should be considered as a principal part.
In Eq. \eqref{eq:10}, $\wk$ is given by the dispersion relation \eqref{eq:12} or \eqref{eq:13} in the effective mass approximation, below and above the gap respectively. In evaluating the integral over $k$ in \eqref{eq:10}, we must pay some attention to the validity of our approximations. This integral should be extended to the full range of $k$ from $0$ to $\infty$. However, if the atomic transition frequency is close to one of the band edges, only a much more restricted range of wavevectors gives a relevant contribution to the integral, due to both the resonance condition and the density of photonic states. In fact, if $E_i \simeq \hbar \omega_c$ with $E_i > \hbar \omega_c$, only modes with $k \simeq k_0$ and $k>k_0$ (that is just above the gap) are relevant, for two reasons: these modes are resonant with the atomic transition frequency, and the density of states is very high for them.
For such modes, the dispersion relation \eqref{eq:13} is a very good approximation. Also, modes with $k \simeq k_0$ and $k < k_0$ (that is just below the gap), which also have a high density of states, are far from resonance, in particular if the gap interval $\omega_c - \omega_v$ is sufficiently large, and thus they are not expected to give an important contribution. We shall therefore neglect them compared to modes just above the gap. All these considerations allow us to use $k_0$ and $\infty$ as, respectively, lower and upper limit of integration in \eqref{eq:10}, and to use the effective mass approximation \eqref{eq:13}. The fact that in this way we are also including modes with wavevector $k$ well above $k_0$, for which the effective mass approximation is not valid, does not introduce a significant error because they give a small contribution to the integral, due to the low density of states and/or the off-resonance condition for these field modes.

The $k$ integral in Eq. \eqref{eq:10} that we need to calculate is thus (assuming $z \simeq \hbar \omega_c$, according to the discussion above)
\begin{eqnarray}
I(z,r) &=& \int_{k_0}^\infty \! dk \frac {\omega_c + A (k-k_0)^2}{z -\hbar [\omega_c + A (k-k_0)^2]}\frac {\sin (kr)}{kr}
\nonumber \\
&=& \int_0^\infty \! du \frac {\omega_c + A u^2}{z -\hbar (\omega_c + A u^2)}\frac {\sin [(u+k_0)r]}{(u+k_o)r} \, ,
\label{eq:16}
\end{eqnarray}
where in the second line the substitution $u=k-k_0$ has been done. This integral can be calculated analytically, yielding
\begin{widetext}
\begin{eqnarray}
I(z,r) &=& \frac 1{2r} \frac 1{\sqrt{\hbar A}\sqrt{z-\hbar \omega_c}}
\left\{ \left(  \frac {\omega_c \pi}
{\sqrt{\frac {z-\hbar \omega_c}{\hbar A}}+k_0} +
\frac {A \pi \frac {z-\hbar \omega_c}{\hbar A}}
{\sqrt{\frac {z-\hbar \omega_c}{\hbar A}}+k_0} \right)
\left[ \sin (k_0r) \sin \left( \sqrt{\frac{z-\hbar \omega_c}{\hbar A}}r \right) \right. \right. \nonumber \\
&-& \left. \left. \cos (k_0r) \cos \left( \sqrt{\frac{z-\hbar \omega_c}{\hbar A}}r \right)
\right] \right\} \, .
\label{eq:17}
\end{eqnarray}
\end{widetext}

Substitution of \eqref{eq:17} into \eqref{eq:10} gives an implicit equation in the variable $z$, whose solution gives the energy shift we are looking for. We first consider an iterative solution of this equation, which is equivalent to a perturbative expansion. First iteration is obtained by replacing, in the right-hand side of Eq. \eqref{eq:10}, the unperturbed value $E_i$ of the energy in place of $z$, obtaining $E=E_i+ \Delta + \delta E$, with $\Delta$ a distance-independent quantity (see Eqs. \eqref{eq:9} and \eqref{eq:10}) and
\begin{equation}
\delta E \simeq \frac \hbar\pi \left( \mu^{eg} \right)_m  \left( \mu^{ge} \right)_n \left( -\nabla^2 \delta_{mn}+\nabla_m \nabla_n \right) I(E_i,r) \, .
\label{eq:18}
\end{equation}

Because of our assumption of an atomic transition frequency close to the edge of the photonic gap, that is $E_i \simeq \hbar \omega_c$, the expression \eqref{eq:17} can be simplified, obtaining
\begin{equation}
I(E_i \simeq \hbar \omega_c,r) = \frac 1{2r} \frac {\pi \omega_c}{\sqrt{\hbar A}\sqrt{E_i-\hbar \omega_c}} \frac {\cos (k_0r)}{k_0r} \, ,
\label{eq:19}
\end{equation}
and substitution into \eqref{eq:18} finally yields
\begin{eqnarray}
\delta E(r) &\simeq& \frac {\hbar \omega_c}{2\sqrt{\hbar A}\sqrt{E_i-\hbar \omega_c}} \left( \mu^{eg} \right)_m  \left( \mu^{ge} \right)_n
\nonumber \\
&\times& \! \! \left( -\nabla^2 \delta_{mn}+\nabla_m \nabla_n \right) \frac {\cos (k_0r)}{k_0r} \, .
\label{eq:20}
\end{eqnarray}
A more explicit expression of \eqref{eq:20} can be easily considered in the long-distance case $r \gg k_0^{-1}$ we are considering. In this case, being
\begin{eqnarray}
& &\left( -\nabla^2 \delta_{mn}+\nabla_m \nabla_n \right) \frac {\cos (k_0r)}r
\nonumber \\
& & \simeq -{k_0^2} \left( \hat{r}_m \hat{r}_n -\delta_{mn} \right)
\frac {\cos (k_0r)}r \, ,
\label{eq:21}
\end{eqnarray}
we obtain
\begin{eqnarray}
\delta E(r) &\simeq& -\frac {\omega_c k_0}{2\sqrt{A(E_i/\hbar -\omega_c )}} \left( \mu^{eg} \right)_m  \left( \mu^{ge} \right)_n
\nonumber \\
&\times& \! \! \left( \hat{r}_m \hat{r}_n -\delta_{mn}\right) \frac {\cos (k_0r)}r \, .
\label{eq:21a}
\end{eqnarray}

Therefore, in the quasi-static approximation the resonance force between the two atoms in the photonic crystal is
\begin{eqnarray}
F_{PC}(r)&=& - \frac \partial{\partial r} \delta E(r) \simeq -\frac {\omega_c k_0^2}{2\sqrt{A(E_i/\hbar -\omega_c )}}  \nonumber \\
&\times& \! \! \left( \mu^{eg} \right)_m  \left( \mu^{ge} \right)_n \left( \hat{r}_m \hat{r}_n -\delta_{mn}\right) \frac {\sin (k_0r)}r
\label{eq:21b}
\end{eqnarray}
($r \gg k_0^{-1}$).

The expression \eqref{eq:21b} of the force should be compared with the analogous expression obtained in the case of two identical atoms, prepared in the entangled state \eqref{eq:2} and in the long-distance approximation (given in this case by $r \gg (E_i/\hbar c)^{-1}$), in the vacuum space \cite{CP98,MP64}
\begin{equation}
F_{vac}(r) \simeq \left( \frac {E_i}{\hbar c} \right)^3 \left( \mu^{eg} \right)_m  \left( \mu^{ge} \right)_n
\left( \hat{r}_m \hat{r}_n -\delta_{mn}\right) \frac {\sin (\frac {E_i}{\hbar c}r)}r \, .
\label{eq:22}
\end{equation}

Comparison of \eqref{eq:21b} with \eqref{eq:22} shows that for atoms in the isotropic photonic crystal, with frequencies close to the upper edge of the gap, the asymptotic behavior of the force as $1/r$ does not change, and that a extra factor $\omega_c/(2\sqrt{A(E_i/\hbar -\omega_c )})$ is present. This extra factor gives an enhancement of the resonant force, particularly relevant when the transition frequency of the atoms is close to the edge of the photonic gap. The force diverges if $E_i = \hbar \omega_c$, due to the van Hove singularity of the density of states \eqref{eq:14}. All this shows how the environment can strongly modify the interaction energy between the atoms and the resonant force between them.

The divergence of the force \eqref{eq:21b} when $E_i \to \hbar \omega_c$ from above, is not physically realistic, of course. In fact, in this limiting cases, higher order terms in the interaction, neglected in our first-iteration solution \eqref{eq:18}, should be included. Higher-order terms take into account that our initial state \eqref{eq:2} should spontaneously (exponentially) decay, and thus this state has a natural width $\Gamma$ ($1/\Gamma$ is the decay time of the state \eqref{eq:2} inside the photonic crystal). We could reasonably consider valid our approximations if $E_i - \hbar \omega_c > \Gamma$. More specifically, a phenomenological inclusion of the effect of the higher-order terms yielding the broadening of our state $\mid i \rangle$, could be obtained by adding an imaginary part, related to its spontaneous decay rate, to the atomic frequency $E_i/\hbar$. This finally gives the following expression for the interatomic force
\begin{eqnarray}
F_{PC}(r)&=& - \frac \partial{\partial r} \delta E(r) \simeq -\frac {\omega_c k_0^2}{2\sqrt{A(\mid E_i/\hbar -\omega_c \mid + \Gamma )}}  \nonumber \\
&\times& \! \! \left( \mu^{eg} \right)_m  \left( \mu^{ge} \right)_n \left( \hat{r}_m \hat{r}_n -\delta_{mn}\right) \frac {\sin (k_0r)}r \, .
\label{eq:23}
\end{eqnarray}
The quantity $\sqrt{A(\mid E_i/\hbar -\omega_c \mid + \Gamma )}$ in the denominator of \eqref{eq:23}, taking into account that $\Gamma >0$ and $E_i > \hbar \omega_c$, cannot vanish.
The maximum increase of the force, relative to the case of atoms in the vacuum space, is now obtained when $E_i=\hbar \omega_c$ and, from \eqref{eq:22} and \eqref{eq:23}, it is given by
\begin{equation}
\left\vert \frac {F_{PC}}{F_{vac}} \right\vert \sim \frac{\omega_c k_0^2/2\sqrt{A\Gamma}}{(E_i/\hbar c)^3} \, .
\label{eq:24}
\end{equation}

Depending on the specific atoms and photonic crystal, the ratio \eqref{eq:24} can be also of some orders of magnitude. For example, taking typical values such as $E_i/\hbar \sim 10^{15} \, \text{s}^{-1}$, $k_0 \sim 10^7 \text{m}^{-1}$, $A \sim 10^2 \, \text{m}^2 \text{s}^{-1}$, $\Gamma \sim 10^2 \Gamma_{Vac}$, where $\Gamma_{Vac} \sim \, 10^8 \text{s}^{-1}$ is the natural frequency broadening of the atomic excited level in the vacuum, from \eqref{eq:24} we obtain $\vert F_{PC}/F_{vac} \vert \sim 10^3$. We wish to mention that analogous results are obtained if the atomic transition frequency is close to the low frequency $\omega_v$ of the gap, using in this case the dispersion relation \eqref{eq:12}. Relations analogous to \eqref{eq:20}, \eqref{eq:21a} and \eqref{eq:23} are then obtained, with $\omega_c$ replaced by $\omega_v$.

The results obtained show that a quite large enhancement of the resonant force is feasible.

It is thus worth to consider the resonant force between the two atoms also using a 1D model for the photonic crystal. In the 1D case, the singularity at the edge of the photonic bandgap is present too, and it is not related to extra assumptions as in the isotropic case.
A realistic realization of this 1D case could be obtained, for example, when the photonic crystal is inside a cavity such that only photons along the $\hat{x}$ direction ($\hat{x}$ is the direction of the 1D crystal) can propagate.
Due to resonance condition, this is necessary only for photons with a wavelength close to the atomic transition wavelength $E_i/\hbar c$.

Our starting point is equation \eqref{eq:10c}. Following the same procedure and approximations used above for the isotropic case, using the 1D dispersion relation \eqref{eq:13} in the effective mass approximation, the $k$ integral in \eqref{eq:10c} after some algebra becomes
\begin{eqnarray}
I(z,x) &=& \int_{k_0}^\infty \! dk \frac {\omega_c + A (k-k_0)^2}{z -\hbar [\omega_c + A (k-k_0)^2]}\cos(k_0 x) \nonumber \\
&=& \int_0^\infty \! du \frac {\omega_c + A u^2}{z -\hbar (\omega_c + A u^2)}\cos((u+k_0)x)\, ,
\label{eq:25}
\end{eqnarray}
This integral can be easily calculated, giving (in the far-zone limit $x \gg k_0^{-1}$)
\begin{widetext}
\begin{eqnarray}
I(z,x) &=&
\left(\frac{\omega_c \pi}{2\hbar A}\frac 1{\sqrt{\frac {z-\hbar \omega_c}{\hbar A}}}+\frac{\pi}{2\hbar} \sqrt{\frac {z-\hbar \omega_c}{\hbar A}} \right)
\left (\cos (k_0x) \sin\left( \sqrt{\frac{z-\hbar \omega_c}{\hbar A}}x \right)
+\sin (k_0 x) \cos \left( \sqrt{\frac{z-\hbar \omega_c}{\hbar A}}x \right)\right)\, .
\label{eq:26}
\end{eqnarray}
\end{widetext}
We now substitute \eqref{eq:26} into \eqref{eq:10c}; iterative solution of \eqref{eq:10c} at first order, finally gives
\begin{eqnarray}
\delta E^{(1D)}\simeq \pi
\frac{\omega_c \left( \mid \bp^{eg} \mid^2 - \mid \bp^{eg}_x \mid^2  \right)}{\sqrt{A(E_i/\hbar-\omega_c)}}\sin(k_0 x) .
\label{eq:26a}
\end{eqnarray}
Therefore the resonant Casimir-Polder force between two atoms in a 1D photonic crystal, in the far-zone limit $(x\gg k_0^{-1})$, is
\begin{eqnarray}
F^{(1D)}_{PC}\simeq - \pi
\frac{\omega_c k_0\left( \mid \bp^{eg} \mid^2 - \mid \bp^{eg}_x \mid^2 \right)}{\sqrt{A(E_i/\hbar-\omega_c)}}\cos(k_0 x) .
\label{eq:27}
\end{eqnarray}

We can now compare our expression \eqref{eq:27} with the analogous expression for the resonant Casimir-Polder force between two entangled identical atoms placed in the vacuum, in a 1D model,
\begin{eqnarray}
F^{(1D)}_{vac}\simeq -2\pi \left( \mid \bp^{eg} \mid^2 - \mid \bp^{eg}_x \mid^2  \right) \left (\frac{E_i}{\hbar c} \right)^{2}\cos\left (\frac{E_i}{\hbar c} x\right)  .
\label{eq:28}
\end{eqnarray}

Eqs. \eqref{eq:27} and \eqref{eq:28} show that the 1D resonant force in both cases, photonic crystal and vacuum space, has the same oscillatory behavior in space (in our approximation $E_i \simeq \hbar c k_0$).
Following the same procedure yielding \eqref{eq:24}, the enhancement of the force with respect to the case of atoms in vacuum, for this 1D case, is easily obtained
\begin{eqnarray}
\left\vert \frac {F^{(1D)}_{PC}}{F^{(1D)}_{vac}} \right\vert \sim \frac{\omega_c k_0/\sqrt{A\Gamma}}{(E_i/\hbar c)^2}  .
\label{eq:29}
\end{eqnarray}

Using the same values for the parameters $A$, $k_0$, $\omega_c$, $E_i/\hbar$ given above, and the atomic decay rate $\Gamma$ in a 1D photonic crystal as obtained in \cite{SH05}, from \eqref{eq:29}
we obtain $\vert F^{(1D)}_{PC}/F^{(1D)}_{vac} \vert\simeq 10$.
This shows that also in the 1D case a significant increment of the resonant force is obtained, when the atomic transition frequency is close to the edge of the band, even if this increase is not as large as in the isotropic case. This confirms that, also in the specific system considered, the isotropic model could overvalue the effect of the crystal, se already suggested in \cite{LLZ00}.

At last, let us compare the present result with the Caimir-Polder force of the one-dimensional electronic system mentioned in the end of Section \ref{sec:2}.
By taking spatial derivative of Eq.(\ref{El:disp}), we can find out that the electronic Casimir-Polder force for the one-dimensional electronic system is proportional to
\begin{equation}\label{PCel}
F_{PC}^e\propto \cos (\kappa_0 x)  \;,
\end{equation}
where $\kappa_0$ is defined by
\begin{equation}\label{InBand}
E_0\equiv -B\cos \kappa_0\;,  \quad \text{ for  } |E_0|\leq B
\end{equation}
(the last inequality in Eq.(\ref{InBand}) means that the impurity energy levels are inside the continuous electronic band). The correspondence between  Eqs.(\ref{eq:27}) and (\ref{PCel}) is obvious, showing a strong similarity of the behavior of the Casimir-Polder force for the two (electromagnetic and electronic) systems.

Finally, we wish to stress that the simplified isotropic 3D model can overvalue the density of states near the band edge because of the singularity, compared to a realistic 3D photonic crystal where numerical simulations have shown that this singularity in the density of states and in the local density of states is absent. This problem does not arise in the one-dimensional case.
Thus radiative effects, such as the increased spontaneous emission rate of atoms embedded in a 3D photonic crystal, could have been overestimated when evaluated using the dispersion relation and density of states given by the 3D isotropic model \cite{LLZ00}, and may lead to false predictions. On the other hand, some predictions of the isotropic 3D model, such as the non-Weisskopf-Wigner decay and the formation of the photon-atom bound state, have been shown to be qualitatively correct \cite{WGWX03}.

\section{\label{sec:4}Concluding remarks}

We have considered the resonant interaction energy and force between two identical atoms, one excited and the other in the ground state, placed in a
photonic crystal. The identical atoms, modeled as two-level systems, are assumed prepared in their symmetrical entangled state and to have the same orientation of the transition dipole moment. This allows us to obtain a closed equation for the resolvent in the limit of a large interatomic separation, where we can neglect virtual intermediate states compared to real ones. From the poles of the resolvent, we can thus obtain an implicit equation for the energy shift of the system due to the atom-field interaction. We have considered two specific cases: a  one-dimensional photonics crystal and a isotropic 3D crystal, where the dispersion relation is assumed valid for any direction of the wavevector.
We have explicitly solved the equation for the resolvent at the first iteration, when the atomic transition frequency is close to the edge of a photonic bandgap, obtaining the potential energy between the atoms in the photonic crystal, and then the force between them in the quasistatic approach.

We have shown that the modified photonic dispersion relation and density of states due to the presence of the photonic crystal, particularly relevant in the proximity of the bandgap edge, yields a strong enhancement of the interatomic resonant force. Although the force obtained by the first iteration diverges if the transition frequency of the atoms coincides with the edge of the photonic band gap, we argue that this divergence can be phenomenologically eliminated by introducing a finite linewidth of the excited states, which derives from higher-order terms. We have also estimated numerically in realistic situations and in the two cases considered (1D model and 3D isotropic model) the enhancement of the resonant force we have found; the results obtained show that the resonant force enhancement can be of considerable size. We have also pointed out that the assumption of a simplified isotropic model of 3D photonic crystal could overvalue the effect. Finally, we have shown that analogous effects occur also in the electronic Casimir-Polder effect of two impurity atoms in a 1D semiconductor wire.

\begin{acknowledgments}
The authors gratefully acknowledge financial support by the Julian Schwinger Foundation, by Ministero dell'Istruzione, dell'Universit\`{a} e della
Ricerca, by Comitato Regionale di Ricerche Nucleari e di Struttura della Materia, and by the MIUR national program PON
R\& C 2007-2013, project Hippocrates Sviluppo di Micro e Nano-Tecnologie e Sistemi Avanzati per la Salute dell'Uomo
(PON02 00355).
\end{acknowledgments}

\end{document}